\documentclass[letterpaper]{article}
\usepackage{aaai}
\usepackage{times}
\usepackage{helvet}
\usepackage{courier}
\usepackage{graphicx}
\usepackage{amsmath}

\title{User Interest and Interaction Structure in Online Forums}
\author{Di Liu, Daniel Percival \and Stephen E. Fienberg \\ Department of Statistics \\ Carnegie Mellon University \\ 5000 Forbes Avenue \\ Pittsburgh, PA 15213 \\ diliu, dperciva, fienberg@stat.cmu.edu}

\begin{document}
\maketitle

\maketitle

\begin{abstract}
We present a new similarity measure tailored to posts in an online forum.  Our measure takes into account all the available information about user interest and interaction --- the content of posts, the threads in the forum, and the author of the posts.  We use this post similarity to build a similarity between users, based on principal coordinate analysis.  This allows easy visualization of the user activity as well.  Similarity between users has numerous applications, such as clustering or classification.  We show that including the author of a post in the post similarity has a smoothing effect on principal coordinate projections. We demonstrate our method on real data drawn from an internal corporate forum, and compare our results to those given by a standard document classification method. We conclude our method gives a more detailed picture of both the local and global network structure.
\end{abstract}

\section{Introduction}

Social network analysis has grown as a topic of interest with the growth of the internet as an interactive environment, especially in connection with online communities.  The general goals of these approaches include characterizing user behaviors and interactions, as well as extracting information from actual user discussions.  In this paper, we define a measure of similarity between users of an online forum, based on a modification of document classification, which takes into account both their interests and interactions.

Establishing a notion of distance or similarity between the people in a social network provides a useful way to illustrate the structure of the social network.  For example, we might define similar people to represent friendship, shared interest, or similarity in skill.  These interpretations give user similarity a wide variety of applications.  For example, recovering friendship from another form of personal interaction data is useful in sociological studies.  People with similar interests could be targeted with a certain advertisement or product suggestion.  A company could assign people with similar skills to work together on a project.

We base our method on establishing a measure of similarity between all posts created by all users of an online forum.  Our measure takes into account both the textual information and the particular context of the online forum.  Usual approaches take only the textual information of the posts into account.  From similarity between posts, we establish  similarity between users.  We then use this similarity to investigate the structure of the social network.


\section{Related Work}

Several previous studies aimed at network structure analysis highlighted areas related to characterizing and clustering users' behaviors, personal qualities, or interests.  For example, in recommendation systems, collaborative filtering works towards this goal, with high profile applications including Netflix~\cite{netflix}, Amazon.com~\cite{amazon} and financial services~\cite{cfilter}.  Other authors use a singular value decomposition~\cite{netflix}~\cite{boltzmann} or variants of $K$-nearest neighbors~\cite{netflix}~\cite{Sarwar00applicationof} to characterize user interest.  

Internet communities such as forums and blogs introduce new challenges for finding patterns of user behaviors. For example, Yang, et al.,  applied social network prestige measures to infer relative expertise of users in a large website in China~\shortcite{taskcn};  Hogg and   Szab\'o examined user behaviors based on activity rates~\shortcite{DBLP:journals/corr/abs-0803-3482};  Holand and Leinhardt measured user behavior based on in- and out-degrees~\shortcite{p1model}.

Relating to our general approach, other studies focused on integrating user and content-based information in social network studies.  Basu, et al.~\shortcite{Basu98recommendationas} proposed an inductive learning approach that uses both collaboritive and content information in predicting user preferences. Taskar et al.~\shortcite{citeulike:3440590} presented a method which makes a connection between a user's personal and network information.  These methods seek to integrate all types of information available in some social networks.


\begin{table*}[t]
  \caption{A summary of the attributes of the  corporate forum data.  Note that the post word counts include stopwords.}
\label{dat-sum}
\vskip 0.1in
\begin{center}
\begin{tabular}{|l|lcccr|}
\hline
\textbf{Attribute} & \textbf{Min} & \textbf{1st Quartile} & \textbf{Median} & \textbf{3rd Quartile} & \textbf{Max}\\
\hline
\textit{Words in a Post}  & 1  & 14 &  28 &  59 & 8980 \\
\textit{Posts made by a User} &1 &   1 &   5  & 22 & 975 \\
\textit{Posts in a Thread} & 1  & 2  & 3  & 5 & 265 \\ 
\hline
\end{tabular}
\end{center}
\vskip -0.1in
\end{table*}

\section{Our Data}

\subsubsection{Online Forums}

We examine user similarity in the context of an online forum.  An online forum is a system designed for the discussion of topics, with each topic separated into its own area, called a thread.  A thread is begun by a user writing a short document, called a post, which introduces the topic or asks a question about the topic.  Typically, this user also writes a separate title for the thread, which summarizes or highlights the thread topic.  Other users can then continue the discussion by adding their own posts to the thread.  Thus each thread in the forum is a place where many users discuss a certain topic.



\subsection{Corporate Forum Data}

Our data come from a global IT company.  The company created an internal forum in order to enhance information flow between employees.  We have data collected from this forum over a one year period from August 2006 until August 2007.  Over this period, 2,974 users wrote 79,128 posts in 20,090 threads.  The users of this forum are skilled IT professionals, and so the topics discussed in this forum are very technical and specific.  The company is interested in grouping employees in creative ways based on the employee's skills, areas of interest, and other strengths.  


By using the available thread ID and user ID information, we can link posts to threads, and authors to posts.   Table~\ref{dat-sum} gives a summary of the attributes of the forum data.  We see that most posts only contain a few words.  As we will see, this makes it difficult to apply traditional document classification methods, which treat each post as a document.  We also see that most users write only a few posts, and each thread is only a few posts in length.  All of this means that most posts have very little or no content, thread, and user information in common.  Our method will seek to address these issues.

\section{Method}

Our method consists of two main steps.  In the first step, we create a matrix which measures the similarity between all pairs of posts in the forum.  In the next step, we build a similarity matrix for users by creating a coordinate system based on the similarity matrix from the first step.  The results of this second step allow us to examine the structure of the relationships and activity of the forum users.

\subsection{Measuring Similarity Between Posts}

If we consider each post as a document, then our goal is to establish a notion of similarity between the documents.  Methods such Latent Dirichlet Allocation (LDA)~\cite{lda1} and cosine similarity~\cite{TextAnalysis} have been shown to be effective document classification techniques.  Since we wish to establish a numerical measure of post similarity, we will modify cosine similarity.

Cosine similarity gives a similarity measure between two documents based on the words within each document.  The simplest approach only considers the words shared by the two documents.  However, cosine similarity also considers the importance of each word.  One way to measure word importance is the tf-idf formula~\cite{TextAnalysis}.  Suppose we have $N$ total documents.  For a word $j$, tf$(D,j)$ represents the frequency of word $j$ in document $D$.  df$(j)$ represents the number of documents containing the word $j$.  tf-idf is defined as:

\begin{align*}
\mbox{tf-idf}(D,j) &= \mbox{tf}(D,j)\times \log_2(N / \mbox{df}(j)).
\end{align*}

The cosine similarity between documents $D_1$ and $D_2$ is defined as:

\begin{align}
\label{norm1}
\mbox{norm}(D_1) &= \sqrt{\sum_j \mbox{tf-idf}(D_1,j)^2}\\
\label{norm2}
\mbox{norm}(D_2) &= \sqrt{\sum_j \mbox{tf-idf}(D_2,j)^2}\\
\label{cosine}
\mbox{cosine}(D_1,D_2) &= \sum_j \left (\frac{\mbox{tf-idf}(D_1,j) \times \mbox{tf-idf}(D_2,j))}{ \mbox{norm}(D_1) \times \mbox{norm}(D_2)} \right).
\end{align}

However, cosine similarity is insufficient for analyzing forum data.  
Cosine similarity is based on word overlap, and is most effective when applied to long documents.  However, a typical forum post is only a few sentences long.  Additionally, cosine similarity mistreats or ignores information available by considering the threads and the author of the posts.  These two issues result in a very sparse similarity matrix --- many documents intuitively related by thread or author have no relation at all.

To address these problems, we modify cosine similarity as follows to take into account all of the available information in forum posts:

\begin{table}[t]
  \caption{A sample thread from the corporate forum data set.  Here the posts in the thread do not share many words in common.  Traditional document classification methods would therefore consider these posts nearly unrelated.}
\label{thread1}
\vskip 0.1in
\begin{center}
\begin{tabular}{|l|p{2.5in}|}
\hline
& \textbf{Thread Title: Madriva 2007 3D desktop}\\
\hline
Post 1 & Anybody tried mandriva 2007? Its cool with a XGL 3D desktop..  But is hungry for RAM..\\
Post 2 & You should give ubuntu 6.10 (or the 7.04 dev) a try. You might also find this interesting: [HYPERLINK]\\
Post 3 & And lookout for KDE Plasma. More info in : [HYPERLINK]\\
Post 4 & Here are few resources on getting Beryl (beryl.. is extremely irresistable.. enter at your own risk :-) )

[HYPERLINK] [HYPERLINK] [HYPERLINK] (best of all)\\
\hline
\end{tabular}
\end{center}
\vskip -0.1in
\end{table}

\begin{itemize}
\item We append to each post the title of the thread in which it appears.  This makes posts within the same thread more similar in word content and therefore closer in cosine similarity.  

Posts made in the same thread might share little or no words in common, even though they are on the same topic.  Such posts would not be considered related under usual cosine similarity.  Table~\ref{thread1} illustrates this problem via an example of a typical thread in our data set.

We use the thread titles since they roughly represent the topic of the thread.  Additionally, a user typically only reads the title of the thread before deciding to read the rest of the thread and then possibly making a response post.  Therefore, the thread title captures both post topic and user interests.

\item We modify the tf-idf$(D,j)$ measure of word importance to take into account the thread in which document $D$ appears. tf-idf measures word importance only using the overall frequency of a word.  However, if a word appears often in a particular thread, then it is likely to be of particular importance to the thread topic, whether or not it is a common word in an overall sense.  Table~\ref{thread2} gives an example of a thread which illustrates this point.

We define $T(D)$ to be the document consisting of the concatenation of all posts in the thread containing document $D$.  We then define:
$$
\mbox{df}_{T(D)}(j) = \frac{\mbox{df}(j)}{\mbox{tf}(T(D),j)}.
$$
Which gives us the following formula:
\begin{align}
\label{threadtfidf}
\mbox{tf-idf}_{T(D)}(D,j) = \mbox{tf}(D,j)\times \log_2(N / \mbox{df}_{T(D)}(j)).
\end{align}
This new measure takes into account the importance of a word within a thread.  Examining the original df$(j)$ measure, we see that as df$(j)$ increases, the importance of word $j$ goes down.  Dividing by the thread word frequency $\mbox{tf}(T(D),j)$ means $\mbox{df}_{T(D)}(j)$ decreases as a word becomes more common within a thread.  

Note that in combination with the previous point, we have that the words in the thread title are of great importance to the thread topic.  Since the thread title usually represents the topic, this is a desirable effect.

\item After computing the cosine similarity using the above modifications, we add an additional term to capture our belief that documents authored by the same user are similar.  Since we want this term to be independent of both post content and the particular user, this term should be a universal constant.

Our goal is not to cluster posts or to assign posts to users.  Rather, we are interested in examining the relationships between users.  This term does not affect the distance of posts written by different users.  Therefore, the inclusion of an author term is not a circular step.  However, this term plays an important role in our analysis which we discuss later.
\end{itemize}

We therefore modify the cosine similarity equations by replacing df$(j)$ with $\mbox{df}_{T(D)}(j)$, and by replacing each post $D$ with post $D^*$, which has the thread title appended.  We define the function $U(D)$ to return the author of post $D$.  We then define:

{\begin{align}
\label{simeq}
\mbox{sim}(D_1,D_2) &= \mbox{cosine}_{T(D)}(D_1^*,D_2^*) + \lambda I_{\{U(D_1) = U(D_2)\}}\\
\label{disteq}
\mbox{dist}(D_1,D_2) &= \max\left(0,1-\mbox{sim}(D_1,D_2)\right).
\end{align}}

Here, $\lambda$ is our universal author constant as discussed above and $\mbox{cosine}_{T(D)}(D_1,D_2)$ represents the cosine distance in equation~\ref{cosine}, with the modified tf-idf measure given in equation~\ref{threadtfidf} replacing tf-idf in both the distance and the norms in equation~\ref{norm1} and~\ref{norm2}. We then convert the similarity measure in equation~\ref{simeq} to a dissimilarity measure via equation~\ref{disteq}.  Note that the maximum of the cosine similarity measure is 1.  This formula is applied to all pairs of posts, giving us a dissimilarity matrix between all posts.

\begin{table}[t]
  \caption{A sample thread from the corporate data set.  Here the words ``data'' and ``migration'' appear frequently in the posts.  Therefore, within this thread, these words should be given high importance.  In the usual tf-idf framework, these words would be given high importance only if they were relatively rare throughout the forum.}
\label{thread2}
\vskip 0.1in
\begin{center}
\begin{tabular}{|l|p{2.5in}|}
\hline
& \textbf{Thread Title: data migration}\\
\hline
Post 1 & Basically what is data migration?\\
Post 2 & Data migration, basically means to porting data from one environment (format/OS/Database/Server etc) to other environment.\\
Post 3 & The process of translating data from one format to another. Data migration is necessary when an organization decides to use new computing systems or database management system that is incompatible with the current system. Typically, data migration is performed by a set of customized programs or scripts that automatically transfer the data.\\
Post 4 & Migrating to higher version also one of the part in data migration.\\
\hline
\end{tabular}
\end{center}
\vskip -0.1in
\end{table}

\subsection{Measuring Similarity Between Users}

We now seek to create a dissimilarity matrix between all users in the forum, given the dissimilarity matrix between all the posts.  We first seek to visualize the relative position of all the posts in some low dimensional space.  Note that our dissimilarity matrix only gives us a function of the position of the posts, it does not give the coordinates directly.  Therefore, in order to visualize this result and to facilitate further computation, we find a low dimensional representation of the posts which preserves the geometry implied by the original dissimilarity matrix.  We achieve this via principal coordinate analysis~\cite{TESL}.

The method proceeds as follows.  Given a dissimilarity matrix $M$, we obtain the singular value decomposition:

\begin{align*}
M = U\Sigma V^T.
\end{align*}

Where $\Sigma$ is a diagonal matrix consisting of the square roots of the eigenvalues of $M$.  The matrices $U$ and $V$ are matrices whose columns are the eigenvectors of $MM^T$ and $M^TM$, respectively.  $\Sigma V^T$ gives a projection of the rows of $M$ into a new coordinate system.  

If we only use the first few coordinates in this projection, this gives us a low dimensional representation of the data, where the distance between all the posts are preserved as best as possible (for more on multidimensional scaling, see~\citeauthor{MDS}~\citeyear{MDS}).  Note that these coordinates are given in order of importance, so we can usually use only a few and capture a great majority of the geometry.  The relative importance of the coordinates is usually deduced from the eigenvalues.

Using this low dimensional representation of the posts, we can easily visualize the relationships and properties of the users.  For example, by plotting the first two principal components we can see the relative position and spread of each user's posts.  This allows us to visually investigate which users are similar to each other, as well as which users post about a wide variety of topics and in a wide variety of threads. 

Towards the main goal of this paper, we can also use this representation to characterize users.  We can give each user a single set of coordinates by finding the centroid of all that user's posts in this low dimensional space.  This roughly gives us a center which follows the areas of high density.

Using these centroids, we can simply make a distance matrix between users by taking the euclidean distance between all pairs of user centroids.  This distance matrix can characterize the social network structure in a wide variety of ways, such as clustering, spanning trees, or nearest neighbor methods.




\subsubsection{The Universal Author Similarity Constant $\lambda$}

\begin{figure}[t]
\includegraphics[width=3in]{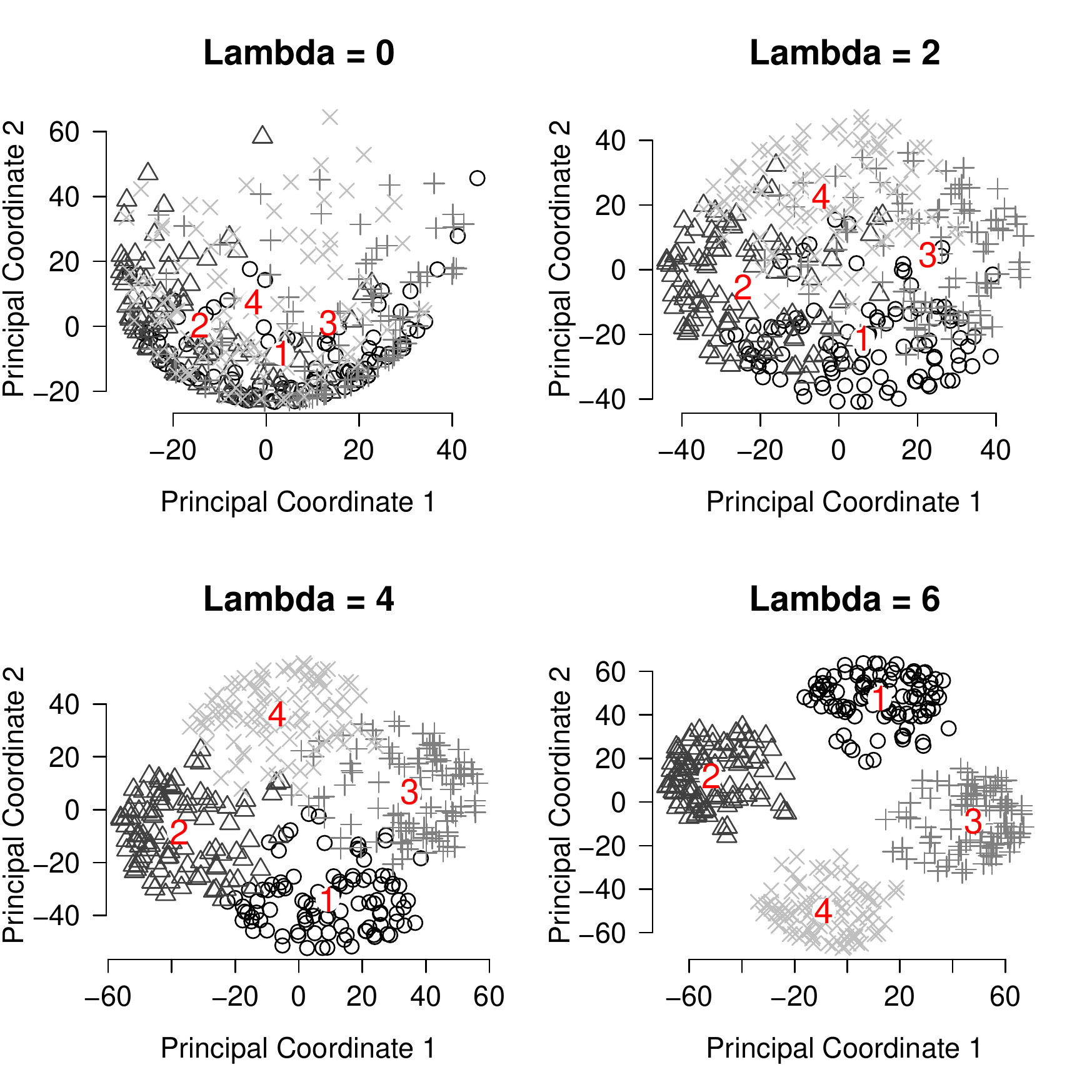}
\caption{Example of the effect of the author constant $\lambda$ on multidimensional scaling.  Here we see that as $\lambda$ increases, the points each of the four groups are drawn towards each other. }
\label{f:sim}
\end{figure}

We now discuss the effects and importance of the author constant $\lambda$.  First we note that $\lambda$ has no effect on pairs of posts written by different authors.  However, it has an effect when we project the posts into a lower dimensional space using principal coordinates.  

Our multidimensional scaling approach attempts to preserve both the distances between pairs of posts written by the same author and pairs of posts written by different authors.  As $\lambda$ grows, posts written by the same author are drawn closer together, while the relative distances between posts written by different authors are held fixed.  The choice of $\lambda$ represents the degree to which we believe an author's posts are similar to each other.  In general, we seek a scaling which preserves and highlights the relationship between authors.

Consider the following example.  We generate four sets of 100 points from different 2-dimensional normal distributions, each with fixed variance $\Sigma = \mbox{diag}(3,3)$, and means $(1,1), (1,-1), (-1,1), (-1,-1)$.  These four groups overlap in the original space, and even after multidimensional scaling it is still difficult to distinguish the groups or characterize their relative positions.  Now consider adding a constant $\lambda$ to the within group similarities.  Figure~\ref{f:sim} shows the effect of increasing $\lambda$ on these data.  The groups are pulled closer together within themselves, but pushed apart from each other.  However, the relative distance and position of the four groups is preserved.  We can think of $\lambda$ as a smoothing term, which reduces within group variance.  $\lambda$ allows a clear visualization and representation of relative group position after multidimensional scaling.

Returning to our application, we recall that we used the projection to estimate the distances between users.  We estimate this with a norm of the difference of two centroids.  As we saw in the example, $\lambda$ has the effect of reducing the variance within each group.  This reduces the variance of the centroid, which is the estimator of the position of the user in this space.  As the norm is a function of the centroids, $\lambda$ reduces the variance of the estimators for the distances between users.  However, $\lambda$ introduces a bias towards a geometry of maximally separated users.  We now discuss this tradeoff in detail.

Suppose we rearrange the rows and columns of the post similarity matrix so each of the $K$ users' posts appear as a block on the diagonal.  Thus, the addition of the author constant term to the dissimilarity matrix can be written as:

\begin{align}
M^* &= M - \lambda H\\
M_{ij} &= 1- \mbox{cosine}_{T(D)}(D_i^*,D_j^*)\\
H &=  \left[
\begin{array}{cccc}
	A_1 & 0 & \ldots & 0 \\
	0  &A_2 &\ldots & 0 \\
	\vdots  &\vdots &\ddots & \vdots \\
	0  &0 &\ldots & A_K \\
\end{array} 
\right] - I.
\end{align}

Where $A_i$ is a square matrix of $1$s with dimension equal to the number of posts written by author $A_i$.  $I$ is the identity matrix with dimension equal to the total number of posts.

We can think of choosing $\lambda$ in terms of two theoretical extremes.  In one case, a very small $\lambda$ has no effect on the scaling.  The Davis-Kahan theorem~\cite{1288832} gives a bound on the difference of the eigenspaces of a matrix $X$, and the matrix $X+Y$.  In our case, since we work with the eigenspace of $M^TM$ we have that this difference is roughly bounded by the Frobenius norm of the matrix $\lambda^2HH^T -2\lambda MH$.  As $\lambda$ becomes small, this norm decreases, and so the author term has a diminishing effect.  The variance of the centroids is not reduced.

In the other case, a large $\lambda$ means that the matrix $M - \lambda H$ is dominated by the second term.  This makes the eigenspace of the matrix $(M- \lambda H)^T(M- \lambda H)$ approach that of $\lambda^2H^TH$.  Since $H$ is block diagonal, then $H^TH$ will also be block diagonal, so the eigenvectors will be roughly piecewise constant.  Consequently, the projection pulls the posts written by each author together into a single point, each of which are maximally separated from the other authors' points.  All the information about relationships between authors has been lost.  $\lambda$ has thus oversmoothed the data.

We therefore choose $\lambda$ small enough to avoid the second case, but large enough so that the information about post author is not completely ignored.  As we argued above, the smoothing effect of $\lambda$ improves the variance of the estimators for user distance.  Pulling together each author's posts somewhat has the desirable effect of separating and clumping authors, thus highlighting their relative positions and controlling outlier documents.

This leads us to choose $\lambda$ based on the nonzero similarities found by applying cosine similarity without considering author, i.e. the nonzero $\mbox{cosine}_{T(D)}(D_1^*,D_2^*)$ terms in equation~\ref{disteq}.  For most forums we recommend using the 75th quantile of these nonzero entires for the value of $\lambda$.  However, for forums with shorter overall thread length, we recommend a higher quantile.  This is because threads give very little information in this case.





\section{Results}

\begin{figure}[t]
\includegraphics[width=3in]{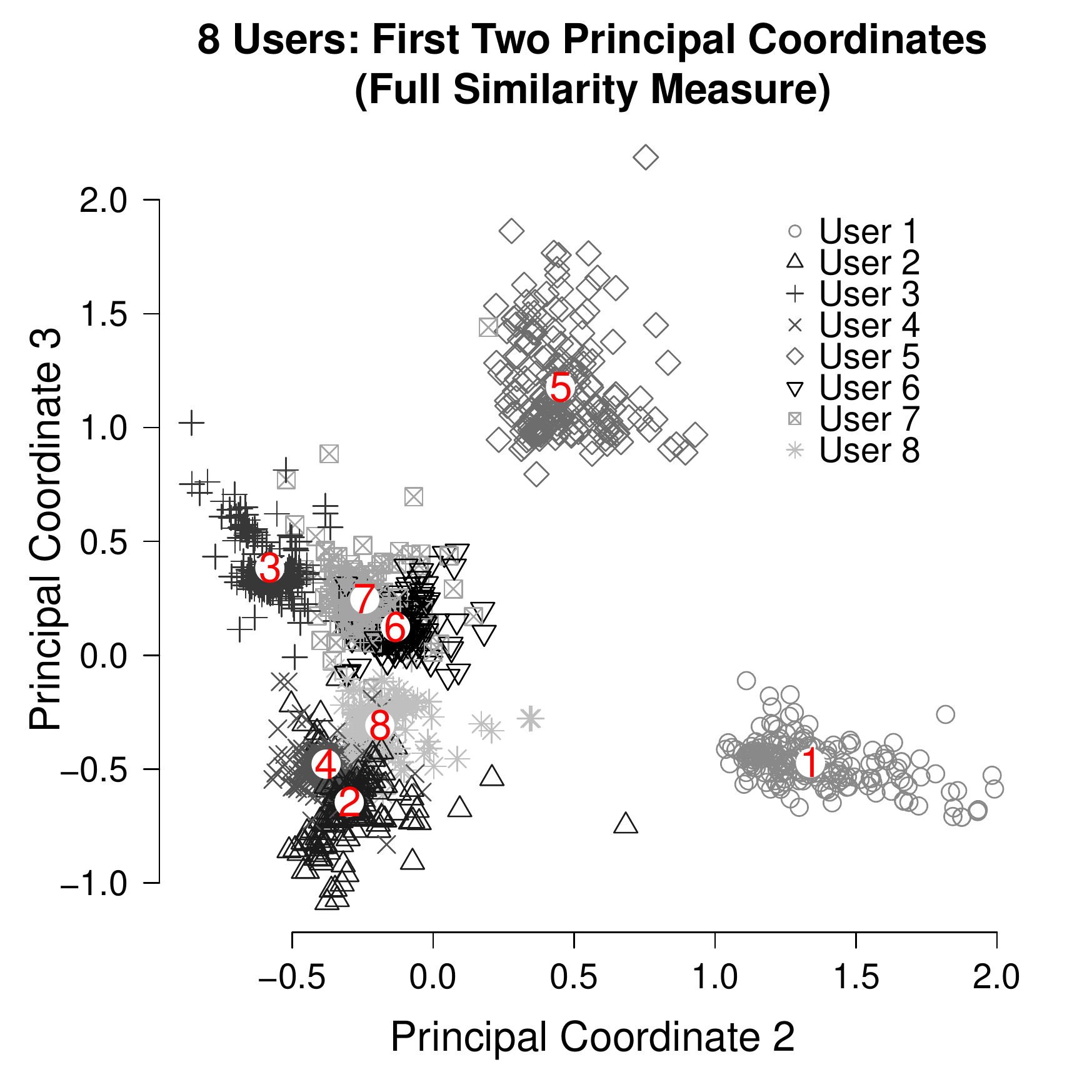}
\caption{The second and third principal coordinates for the 8 user data set.  The centroids are given by the numbers in the white circle.  We can clearly see each user separately, but are still able to see spread information and the similarity between users.}
\label{f:allsim}
\end{figure}

We present results on two subsets of our data.  First, we examine a small set of fairly active users.  This illustrates and visualizes the results of our method.  Second, we present results on a larger set of active users, and evaluate the method via a comparison to document classification.

\subsection{Preprocessing}

We first need to  preprocess the forum post data.  For our method, we first append the thread title to each post.  We next follow conventional text analysis methods, by removing HTML code, removing stopwords, and performing word stemming.  See~\citeauthor{TextAnalysis} for a complete discussion of these standard techniques.



We build a dictionary based on all the words in all of the processed posts.  Thus, each post is now a vector of counts of each word in the dictionary.

\subsection{Illustration: A Small Set of Active Users}

\begin{figure}[t]
\includegraphics[width=3in]{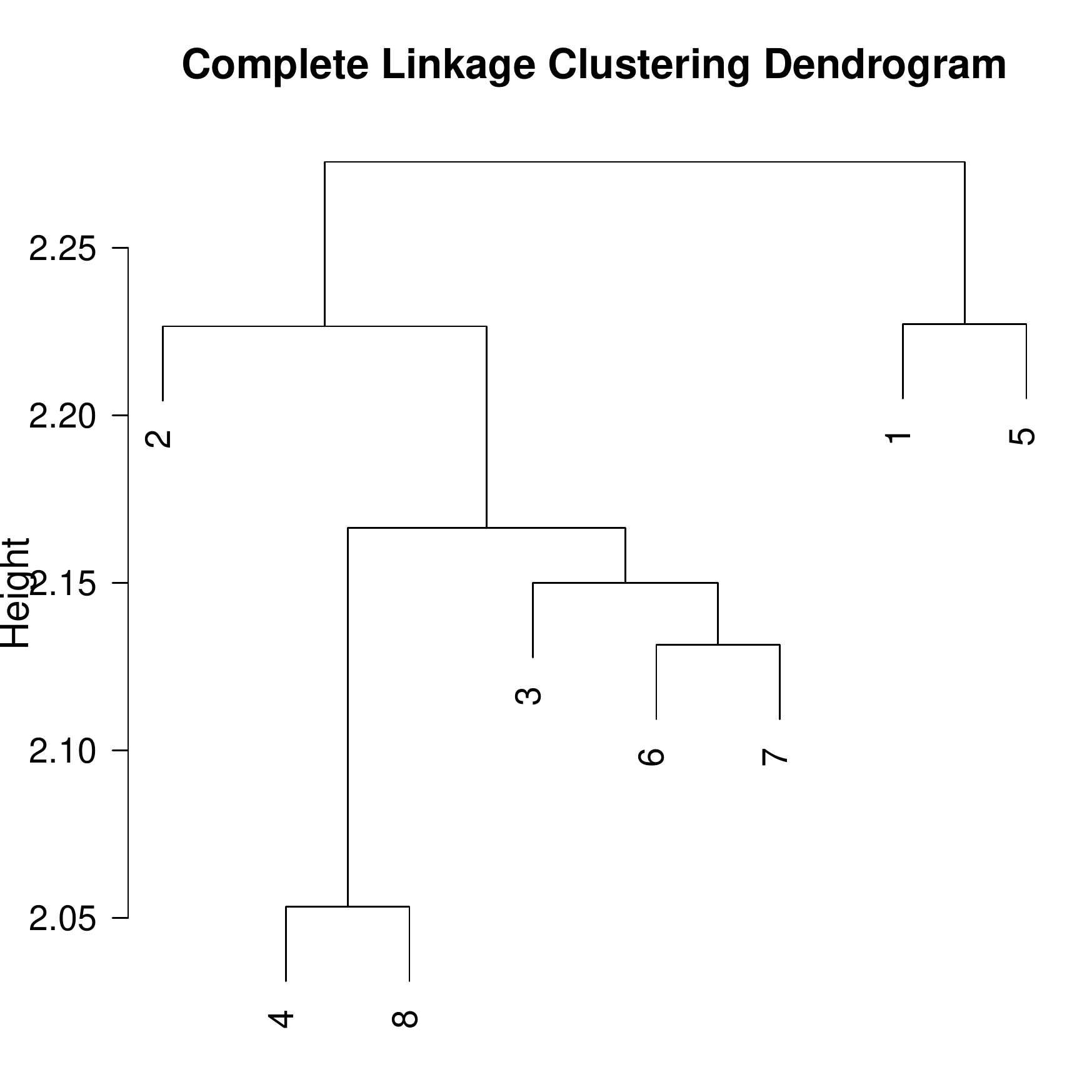}
\caption{Cluster dendrogram for the 8 users.  We can create a clustering by drawing a horizontal line at any height in the tree, and taking the clustering given by the links below the line.  Overall, we see that two clusters seems most appropriate, with clusters $\{1,5\}$ and $\{2,3,4,6,7,8\}$}
\label{f:8utree}
\end{figure}

\begin{figure}[t]
\includegraphics[width=3in]{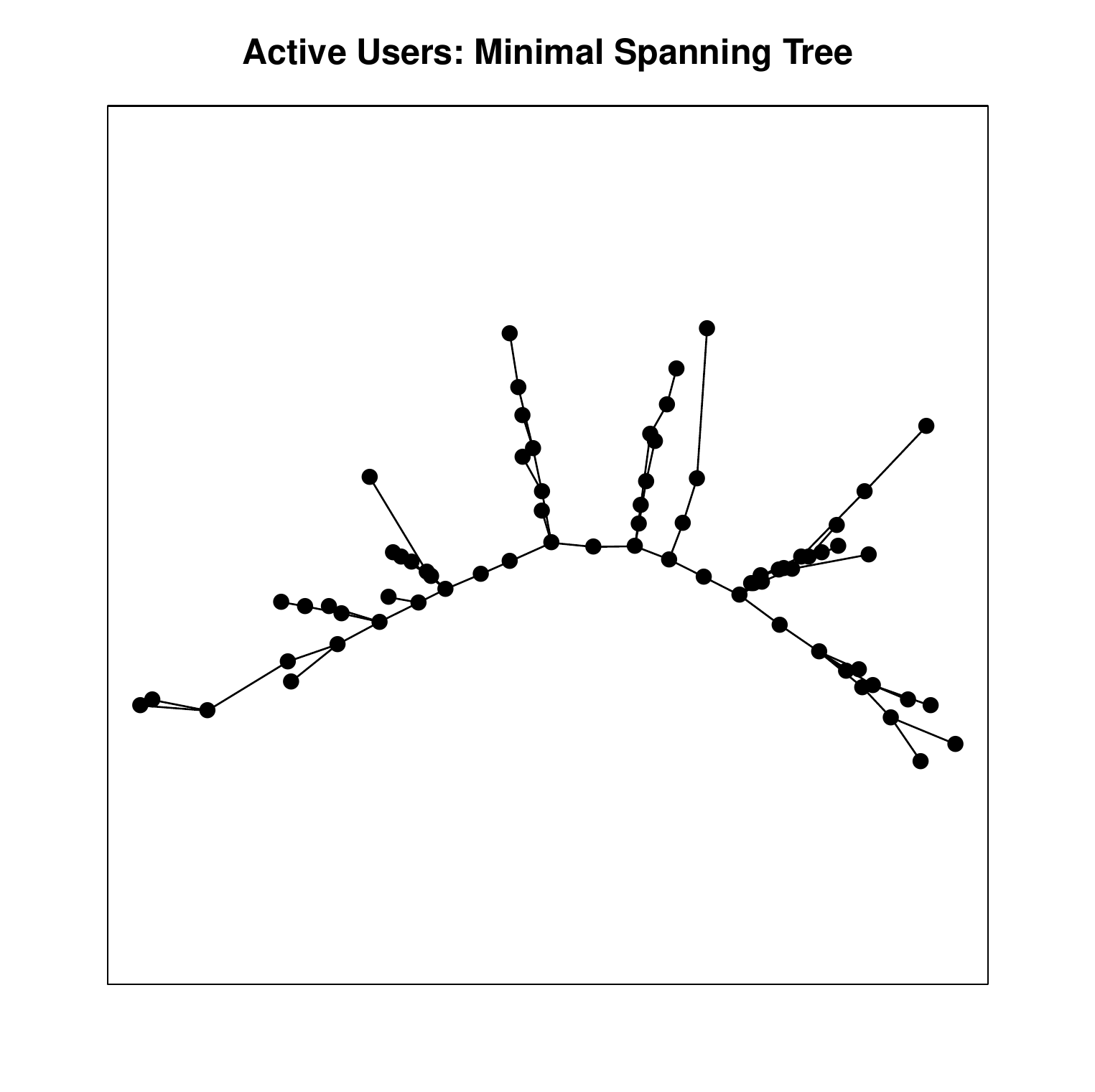}
\caption{The minimal spanning tree for the 71 users.  This tree gives relationships similar to K-nearest-neighbors.}
\label{f:mst}
\end{figure}

In order to illustrate our method clearly and intuitively, we present results for a small set of users.  We consider all users who wrote between 200 and 210 posts on the forum.  In the full dataset, the set of users who wrote more than 200 posts accounts for about 50\% of all posts.  Therefore, this range represents users who are roughly in the middle in terms of posting activity.  These users are also easier to compare since they wrote roughly the same number of posts.  This range gives us eight users in total.

Using only the subset of posts authored by these users, we apply our previously described method.  Note that the dictionary is built from all of the posts, so the tf-idf measures take into account the overall importance of the words.  Therefore, the tf-idf measures are not biased in this case, and so our results for these particular users will not differ greatly from those obtained when all of the posts are included in the analysis.

Figure~\ref{f:allsim} shows the second and third principal coordinates for the posts.  We use $\lambda = 0.059$ for our universal author similarity constant.  This is the $75$th quantile of our nonzero similarities obtained without taking author into account.  We can see a clear separation between users, as well as differing spreads for each user.  Looking at the centroids, we see six users who are somewhat similar, and two users who are separated by these principal coordinates.  These two users also seem to have larger spreads than the other six, which possibly indicates a broader interest in topics.



To illustrate an application of the user distance, we next build a hierarchical clustering tree using complete linkage~\cite{TESL}.  Figure~\ref{f:8utree} displays this tree.  We see that there are two main clusters: users $\{1,5\}$ and users $\{2,3,4,6,7,8\}$.  This is consistent with our earlier display of the users in Figure~\ref{f:allsim}.  This clustering gives a picture of the network structure within this 8-user group.

\subsection{Active Users}

We now consider a large set of active users in the corporate forum data set.  We take all users who wrote between 200 and 400 posts on the forums.  This gives us 71 users and 18,682 total posts.  We consider this subset for computational and interpretive reasons.

We apply our method to these 71 users, with $\lambda = .054$.  This $\lambda$ is obtained from the 75th quantile of the nonzero similarity matrix entries for the modified cosine similarity measure.  Due to the large number of users in this data set, plots of the principal coordinates do not give clear pictures of the relationships between users.  Note that plots of subsets of users can show individual user spread.

As mentioned before, there are many ways to look at the social network structure once we calculate the distance between all of the users.  We present two well-known examples here: complete linkage clustering and a minimal spanning tree~\cite{mst}.

\begin{figure}[t]
\includegraphics[width=3in]{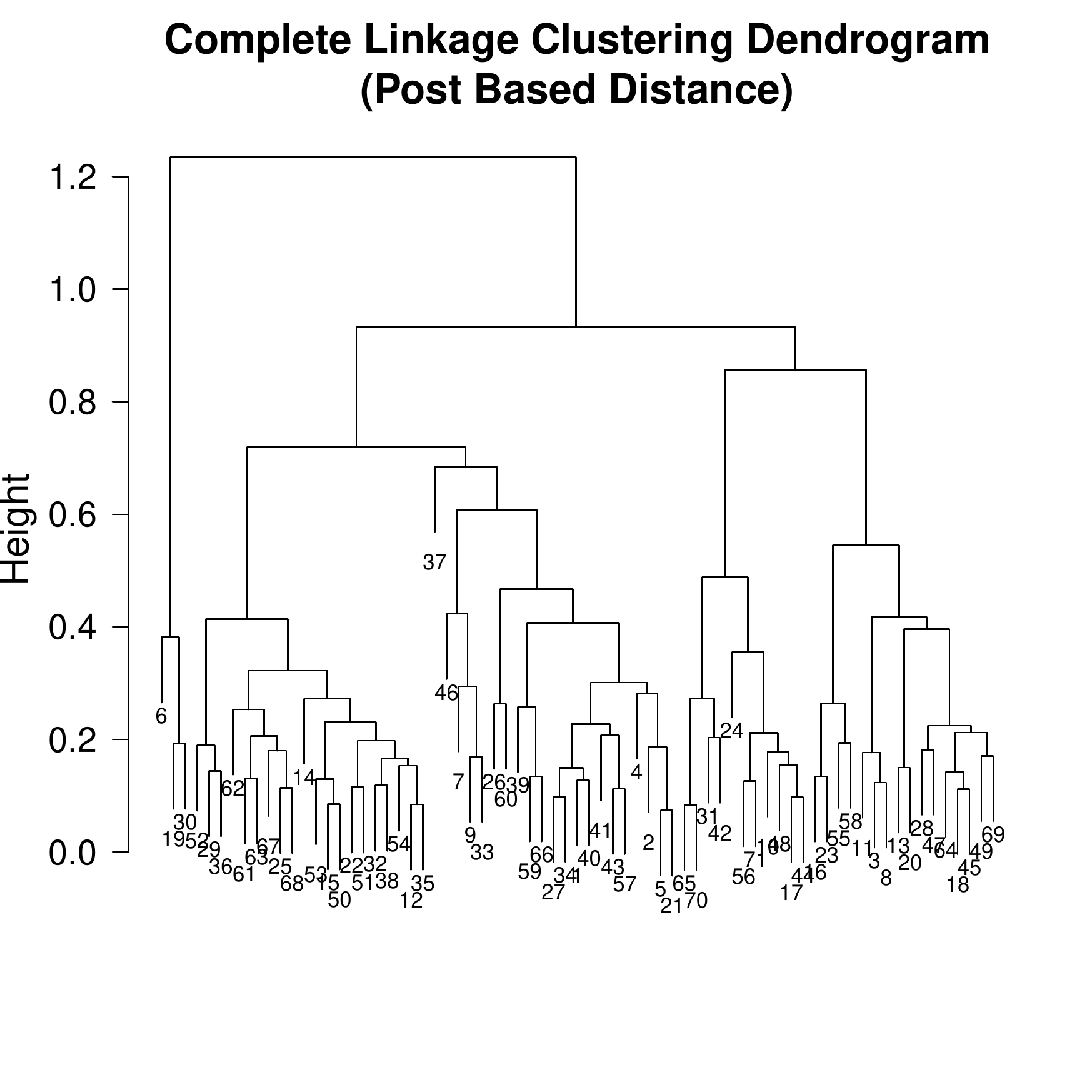}
\caption{Cluster dendrogram for the 71 users.  We can create a clustering by drawing a horizontal line at any height in the tree, and taking the clustering given by the links below the line.  Overall, we see that five clusters seems most appropriate.}
\label{f:71utree}
\end{figure}

Figure~\ref{f:71utree} shows the complete linkage hierarchical cluster dendrogram for the 71 users.  We see that a five cluster solution looks appropriate.  The majority of the users are in two of these five clusters.  The remaining three groups are small and separated from these two large groups.  In particular, we see a group of three users located far away from the other four groups.  This group may represent a collection of users who have the same specialized interest.  The two large groups perhaps deal with general or popular topics.

Figure~\ref{f:mst} displays the minimal spanning tree (MST) for these 71 users.  A minimal spanning tree provides a way to relate users similar to K-nearest-neighbor methods.  Roughly, the closer users are, the more similar they are in terms of interest and interaction.  The minimal spanning tree shows that the structure of the network is varied.  We see several clumps of users in the MST, as well as several users who are very far from any others.  The company could infer that the clumped users share similar interests and often interact, and thus may make a good project team.

\subsubsection{Evaluating the Method}

Since characterizing a social network's structure is not a prediction problem, evaluation of these results can instead be done by comparison to another method with similar output.  For comparison, we approach the problem from a purely conventional document classification viewpoint.  We create a single document for each of the 71 users in the large set by concatenating all of that user's posts (note we do not include thread titles in the posts).  We then use cosine similarity to compare these documents, and thus arrive at a similarity matrix between users.  This similarity tries to measure the relationship between user interest purely based on textual information; it ignores all of the forum structure, such as the threads.

\begin{figure}[t]
\includegraphics[width=3in]{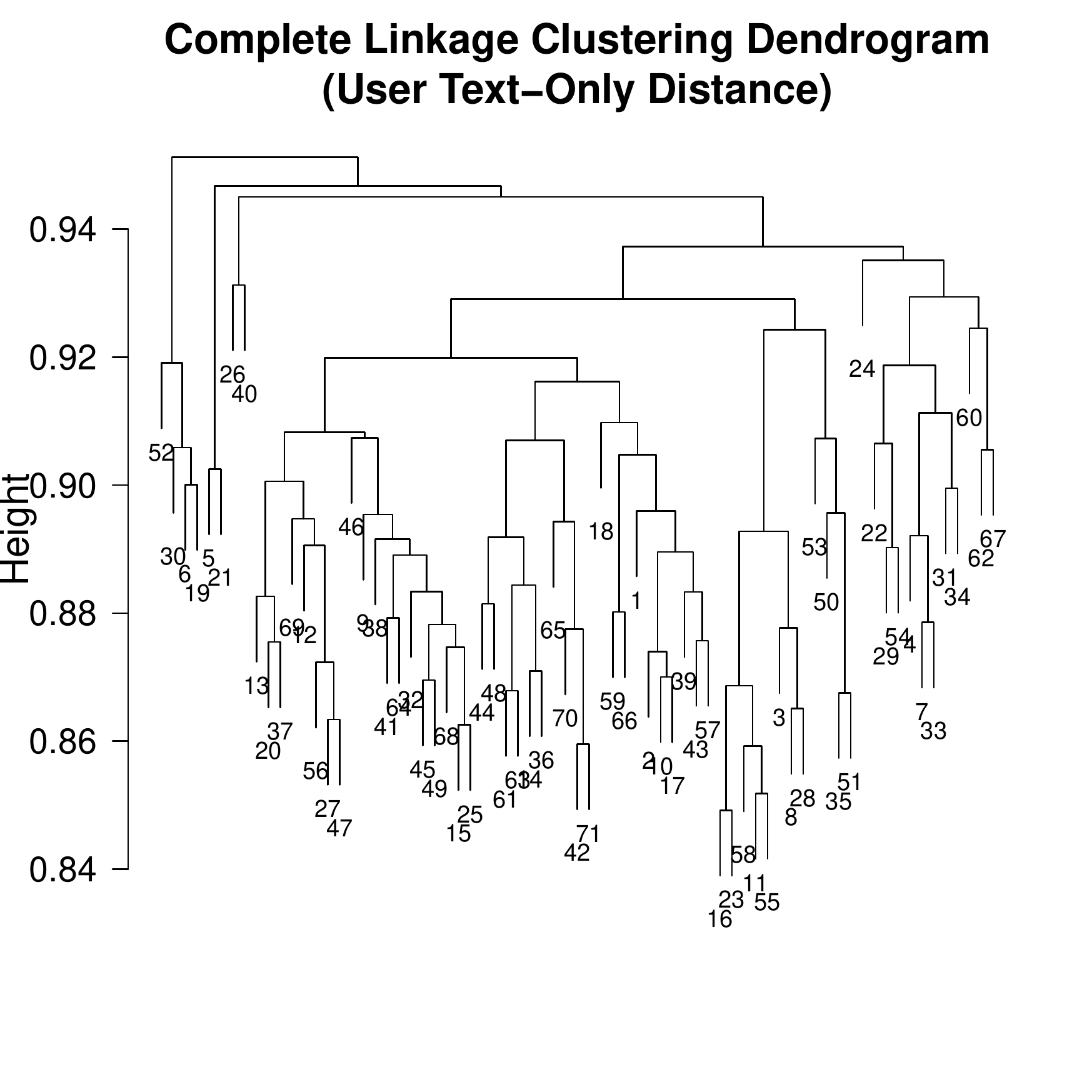}
\caption{Complete linkage clustering using a text-only similarity measure.  There are some similarities to the clustering using our similarity measure, but this tree is not as useful for clustering users.}
\label{f:compclink}
\end{figure}

To be consistent, we also apply complete linkage hierarchical clustering using this similarity measure (see Figure~\ref{f:compclink}).  We see that in both cases, users 30, 6, and 19 are in a cluster which splits from the rest at a large height.  This indicates that both methods found that these users are very far away from the others, due to their different textual information.  We also see that both methods find many pairs of users who are merged into the same cluster at the bottom of the tree.  This indicates that both methods find some similar nearest neighbor pairs.  These similarities show some consistency between the two methods.  Although our method includes a great deal more information than the text of the posts, we still see that the text plays a strong role in defining user similarity.

However, there are differences.  For example, in the pure text based method, user 52 is considered to be in the same group as the ``strange'' users 30, 6, and 19.  On the other hand, our method finds user 52 to be closely related to a different group of users.  Looking at his posts, it is clear that this user often replies to others by posting a hyperlink or an attachment.  Such replies do not contain any text with reference to the topic.  Our method includes the thread titles and therefore links this user to the other users who post in the same threads.  Therefore, out method has captured this user's interactions and interests more fully.

\begin{figure}[t]
\includegraphics[width=3in]{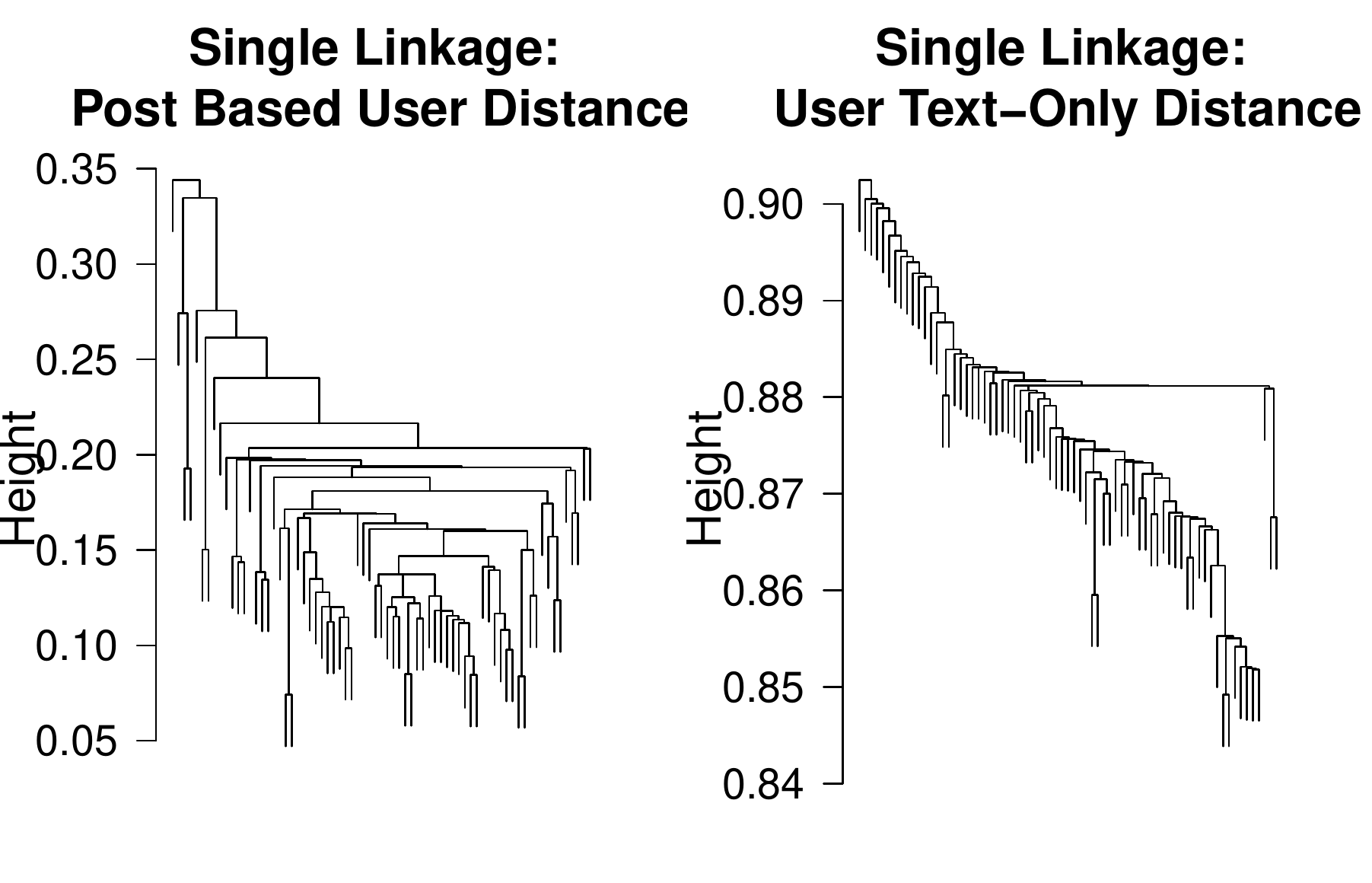}
\caption{Left: single linkage dendrogram using our user similarity.  Right: single linkage dendrogram using the text-only user similarity.   Our method shows more local group structure.}
\label{f:compslink}
\end{figure}

Overall, we see the text-only approach does not appear to have as clear of a group structure.  If we compare the five-cluster solutions in the complete linkage dendrograms, we see that the text-only approach gives four very small clusters, and one giant cluster.  This is not an informative picture of the social network structure.  Complete linkage clustering relies on a global criteria.  We now examine single linkage clustering (refer to figure~\ref{f:compslink}), which relies on local effects.  We see that the text-only approach gives a degenerate structure.  This ``chaining effect'' makes any cluster structure impossible to recover.  On the other hand, our method gives a single linkage structure which still has cluster information.  Therefore, our method gives a more detailed network structure both locally and globally.

\section{Discussion}

Our main contribution is a new similarity measure between posts in a forum.  This measure effectively modifies document similarity to incorporate the special structure of forums.  We discussed the properties of our modification, and presented some results on a real data set. 

Users in a forum demonstrate their interests and interactions with other users in two ways.  First, users write posts whose words can tell us in which topics they are interested.  Second, users post in particular threads, indicating both topic interest and interaction with the other users who have already posted in the thread.  By including the title of threads in each post, we view both types of information in a single unified context.  User interactions within a thread are transformed into shared words.

In our database, thread titles tend to be very short once we  remove stopwords.  Therefore, the longer the post, the smaller the effect of the thread title on cosine similarity.  Since longer posts contain more textual information, this is a desirable effect.  Our modification  give us information about posts which are otherwise hard to characterize.

In our analysis of the corporate forum data, we compared our method to a traditional document classification approach.  We showed that the addition of information about threads and authors was critical for giving a complete picture of the network structure.  This is because posts often do not, in their text, contain references to the topic or to the other users in the thread.  The additional information allows more varied relationships between users.

There are alternatives to the principal coordinates method for defining user dissimilarity.  Given post similarity, we could choose among many methods to obtain user dissimilarity or distance.  For example, we could use the average distance between all pairs of posts written by the two authors:

\begin{align*}
\mbox{dist}(A_1,A_2) = \frac{\sum_{DOC(A_1)} \sum_{DOC(A_2)} \mbox{dist}(D_i,D_j)}{ \left|DOC(A_1)\right|\left|DOC(A_2)\right|},
\end{align*}

where:

\begin{align*}
DOC(A) = \{D: U(D) = A\}
\end{align*}

For this approach, the author term is never included and therefore has no effect.  The principal coordinate approach, however,  seeks to preserve the distance for documents both between and within a user.  Further, $\lambda$ has a smoothing effect on the principal coordinate projections, and if properly chosen improves the variances of the estimates of user distances.  Therefore, our user distances take more information into account.  Principal coordinates also allow for easy visualization of the post and user relationships.

Principal coordinates also can allow additional users to be included in the analysis via a projection.  This allows us to use only a subset of posts or users to generate the coordinates.  We can project the remaining posts into this coordinate space, and thus learn about a larger set of posts or users.  For example, in our data we could use the large set of active users to define coordinates for the great number of low activity users.

We need a more systematic way to pick the author constant $\lambda$.  We believe $\lambda$ also has beneficial properties with regard to statistical testing.  We would like to develop a framework to investigate these properties.  Different estimators of user location besides the centroids proposed in this paper may lead to more rich estimators of user distance.  We are also interested in additional validation of this method on data with some known and recoverable social structure.

\bibliographystyle{aaai}

\end{document}